\documentclass[12pt]{article}

\usepackage[round]{natbib}
\usepackage{setspace}
\usepackage{geometry}
\usepackage[section]{placeins}
\usepackage[hidelinks]{hyperref}
\usepackage{graphicx}
\usepackage{xcolor}
\usepackage{titlesec}
\usepackage[page]{appendix}
\usepackage{enumerate}
\usepackage{epigraph}
\usepackage{lipsum}

\usepackage{ulem}
\usepackage{gitinfo2}

\usepackage{lscape}
\usepackage{booktabs}
\usepackage{rotating}
\usepackage{multirow}
\usepackage{longtable}
\usepackage{caption}
\usepackage{subcaption}
\usepackage{float}
\usepackage{tabularx}
\usepackage{ragged2e}
\newcolumntype{Y}{>{\RaggedRight\arraybackslash}X}
\usepackage{pdflscape}
\usepackage{afterpage}

\usepackage{amsmath}
\usepackage{amssymb}
\usepackage{amsthm}
\usepackage{mathtools}
\usepackage{dsfont}

\usepackage[ruled,vlined]{algorithm2e}

\SetCommentSty{algcommstyle}

\newcommand*{\E}{\mathbb E}

\newcommand*{\I}{\mathds{1}}
\newcommand*{\Prob}{\mathbb P}
\newcommand{\indep}{\perp \!\!\! \perp}
\DeclareMathOperator{\Var}{Var}

\newcommand{\defeq}{\overset{\text{\tiny def}}{=}}

\begin{document}

\title{Approaches to Statistical Efficiency when comparing the embedded adaptive interventions in a SMART}

\author{Timothy Lycurgus, Amy Kilbourne \& Daniel Almirall}

\date{}

\maketitle


\newpage

\abstract{
Sequential, multiple assignment randomized trials (SMARTs), which assist in the optimization of adaptive interventions, are growing in popularity in education and behavioral sciences. 
This is unsurprising, as adaptive interventions reflect the sequential, tailored nature of learning in a classroom or school. 
Nonetheless, as is true elsewhere in education research, observed effect sizes in education-based SMARTs are frequently small. 
As a consequence, statistical efficiency is of paramount importance in their analysis. The contributions of this manuscript are two-fold. 
First, we provide an overview of adaptive interventions and SMART designs for researchers in education science.
Second, we propose four techniques that have the potential to improve statistical efficiency in the analysis of SMARTs. 
We demonstrate the benefits of these techniques in SMART settings both through the analysis of a SMART designed to optimize an adaptive intervention for increasing cognitive behavioral therapy delivery in school settings and through a comprehensive simulation study. 
Each of the proposed techniques is easily implementable, either with over-the-counter statistical software or through R code provided in an online supplement.
}


\newpage

\tableofcontents

\newpage

\section{Introduction}
\label{sec:intro}

In educational settings, individuals or organizations (schools, classrooms, etc.) are often best served by an intervention that is adapted over sequential stages to suit their initial and changing needs.
The salience of an adaptive intervention is, perhaps, most clear in the classroom. 
Conceptual models for learning, themselves, often point toward a sequential, scaffolding approach whereby mastering a given concept frequently necessitates a thorough understanding of the preceding concepts \citep{maybin1992scaffolding}.  
Following an initial lesson or assignment, a classroom teacher may monitor each student to identify those meeting or failing to meet criteria for early signs of success, and then offer each student targeted support based on their needs \citep{arendale1994understanding,rowan2019summary}.
Outside the classroom, as well, there are myriad scenarios where it may be necessary to adapt and re-adapt intervention.  
School principals may need to adjust classroom- or teacher-level interventions 
(e.g., professional development interventions, \citep{bergdahl2022adaptive}) to suit the changing needs of teachers or classrooms. 
Similarly, school districts may need to adjust school-level interventions (e.g., policy interventions designed to improve the adoption of evidence-based practices at schools, \citep{heppen2020can}).

Increasingly, there is interest by educators and education researchers alike in informing how best to make sequences of intervention decisions \citep{raudenbush2008advancing}. 
For example, in Adaptive School-based Implementation of CBT (ASIC), researchers aimed to determine the sequence of interventions that will best improve delivery of cognitive behavioral therapy (CBT) to students within schools \citep{kilbourne2018adaptive}.
Such ``adaptive interventions'', or pre-specified sets of decision rules as to how an intervention should best proceed, guide which treatment should be offered to a student or participant at any given stage of the intervention. 
Also referred to as dynamic treatment regimens or dynamic instructional regimes \citep{raudenbush2008advancing},  these adaptive interventions tailor the provision of treatment to best-serve the changing needs of the participants. 
For example, in the context of one of the adaptive interventions considered in the ASIC Study: If CBT skills coaching for all school professionals at a school does not lead to short-term improvements in CBT delivery for a given school, coaching is augmented with an additional intervention. 
On the other hand, schools that do improve CBT delivery may not need that augmentation.

In some cases, there may be evidence from prior studies, practical expertise, or one or more supporting theories of change that can be used to inform the construction of a high-quality adaptive intervention. 
Here, an education scientist may be happy to proceed with a standard 2-arm confirmatory randomized trial to evaluate the effectiveness of the adaptive intervention versus a suitable control.
In other cases, however, we expect education scientists will have myriad scientific questions that are necessary to answer in order to develop a high-quality adaptive intervention.  
Such questions may include: ``What is the best treatment to offer in the first stage of the adaptive intervention?'', ``How best should we monitor response/non-response to first-stage treatment in a way that is most informative for making second- or subsequent-stage decisions?'', ``At what time points, should a transition to subsequent treatment be considered?'',  ``What second- or subsequent-stage intervention option is best for those who are not responding adequately to prior stage treatment(s)?''

To answer such optimization questions, education researchers may turn to sequential, multiple assignment randomized trials, or SMARTs \citep{lavori2004dynamic,murphy2005experimental}. 
SMARTs are a type of factorial design \citep{murphy2009screening} where some or all participants are randomized multiple times to one or more treatment options, at critical decision points in an adaptive intervention \citep{almirall2018developing}.

SMARTs, frequently utilized in the medical and behavioral intervention sciences, are growing in popularity in education sciences. 
Some of these studies focus on constructing adaptive interventions that directly target skills like reading or math.
For example, \citet{kim2019using} and \citet{fleury2021early} use SMARTs to inform development of adaptive interventions aimed, respectively, at personalizing print and digital content for early elementary students and at improving reading in preschool children with autism.
Other education-based SMARTs focus on constructing adaptive interventions that target learning outcomes indirectly.  
For instance, \citet{pelham2016treatment} use a SMART to determine the appropriate course of action to treat childhood ADHD in the classroom. 

In education science, there exist various frameworks that offer motivation for the use of adaptive interventions in education practice. 
These include, among others,  response to intervention \citep{fuchs2008response} and multi-tiered systems of supports \citep{roberts2021multitiered}.

The primary contribution of this manuscript is two-fold: 
First, we introduce applied statisticians and methodologists in education sciences to a longitudinal data analysis method that can be used to address three of the most common primary aims in a SMART.
Second, and more interestingly, we provide education scientists with a suite of easy-to-implement techniques that, in many cases, can lead to increased statistical efficiency (e.g., narrower confidence intervals or greater statistical power).
The latter, in particular, is especially important in education and other behavioral intervention sciences, where effect sizes for the comparison of adaptive interventions (or the components of an adaptive intervention) are expected to be small to moderate \citep{kraft2020interpreting}.

We illustrate the methods using data from ASIC, a SMART designed to optimize an adaptive implementation intervention to improve mental health interventions in schools \citep{kilbourne2018adaptive}.
Cognitive behavioral therapy has been shown to improve outomes among those affected by depressive and anxiety disorders, but barriers to obtaining CBT limit access among those who are affected.
ASIC compares the effectiveness of a set of adaptive interventions employing various strategies to address barriers to CBT delivery. 
We leverage data from ASIC to illustrate how different techniques, either implemented alone or in tandem with others, may improve efficiency when analyzing SMARTs. 
To demonstrate the benefits of these techniques on statistical efficiency under various scenarios, we present a simulation study implementing these methods on synthetic SMART data. 

We begin by providing a brief introduction to adaptive interventions and sequential multiple assignment randomized trials in Section~\ref{sec:rev}. 
In Section~\ref{sec:eff_methods}, we discuss how to improve efficiency in SMARTs with over the counter methods.
We implement these methods on the ASIC data in Section~\ref{sec:ASIC_res}. 
Section~\ref{sec:sims} demonstrates the benefits of the efficiency techniques presented in Section~\ref{sec:eff_methods} through a comprehensive simulation study.
We conclude with a discussion of the efficiency benefits of these methods and how the work may be extended to more complex SMART designs. 

\section{Review: Adaptive Interventions and SMARTs}
\label{sec:rev}

An adaptive intervention (AI) is a pre-specified set of decision rules that guides how best to serve the needs of individuals from a pre-specified population. 
These rules tailor the provision of treatment at critical decision points during intervention. 
Specifically, there are four aspects of adaptive interventions: decision points, treatment options, decision rules, and tailoring variables \citep{Seewald2020}. 
\textit{Decision points} are the times at which an intervention decision is made; we refer to the set of treatments available at a decision point as the \textit{treatment options}.
Treatment options may include, among others, the type of treatment, the intensity of the treatment, or a combination of two or more individual treatments.
The \textit{decision rule} guides which treatment to select for an individual at a given decision point. 
The decision rule makes this determination based on the value of one or more \textit{tailoring variables}.  
A tailoring variable can be constructed from any known information collected prior to or at the current decision point.
Note that this includes information that could have been impacted by interventions offered at prior decision points. 
For example, a tailoring variable may include static information (e.g., school district or race), or time-varying information (e.g., improvements in academic performance since the prior decision point).

In some cases, researchers may use any one (or a combination) of the following to inform the construction of adaptive interventions: education practice expertise, existing theories of change or conceptual models or frameworks, or evidence from prior studies or observational study analyses including evidence from prior randomized trials.
If there is evidence from prior studies suggesting a given AI will be successful, a standard two-arm randomized trial may be conducted to evaluate the effectiveness of the AI in comparison to the control. 
For such an example, see \citet{raudenbush2020longitudinally}.
Alternatively, researchers may have numerous questions they want to answer in order to construct a more effective adaptive intervention.
We call these optimization questions because their ultimate goal is to generate evidence for a more optimized AI \citep{collins2007multiphase}.
Such optimization questions include: ``Which treatment option should be offered in the first stage of an adaptive intervention?'' or ``What subsequent intervention should be offered to schoolchildren who respond unfavorably to the prior treatment?''
To answer these optimization questions, researchers may use a sequential, multiple assignment randomized trial (SMART). 

In SMARTs, participants take part in multiple stages of the intervention, where each stage corresponds to a decision point where individuals may be randomized to two or more intervention options.
SMARTs stand in contrast to the single-stage-at-a-time experimental approach where a separate randomized trial is conducted and analyzed for each stage of the adaptive intervention \citep{murphy2007developing,nahum2012experimental}. 

\begin{figure}[htb]
  \begin{center}
    \includegraphics[scale = 0.95]{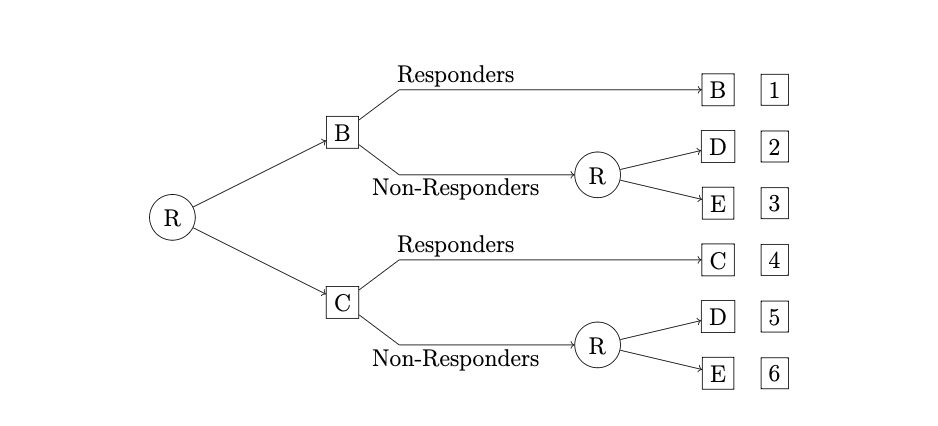}
  \end{center}
\caption{A two-stage prototypical SMART. Circled R denotes a randomization point. B, C, D, and E denote different treatments and 1-6 denote different treatment pathways. For example, ``2'' denotes receiving intervention ``B'' in the first stage and then receiving ``D'' in the second stage.}
    \label{fig:proto_smart}
\end{figure}

There are many different SMART designs, but we focus on the prototypical SMART, seen in Figure \ref{fig:proto_smart}, for the purposes of this paper.
For other common SMART designs, see \citet{almirall2018experimental}. 
In the prototypical SMART, all participants are randomized during the first stage of the treatment.
At subsequent stages, only non-responders are re-randomized to an adjusted treatment. 
In the prototypical SMART, Response/Non-Response is the tailoring variable, i.e., the variable that defines the decision rule such that the treatment is individualized for responders versus non-responders. 
For simplicity, in this paper all randomizations occur with probability 0.5 but in practice, randomization probabilities may vary. 

\subsection{A Common Primary Aim in a SMART}
\label{subsec:primary_aims}

Let $Y_i^{(a_1, a_{2R}, a_{2NR})}$ denote the continuous primary outcome for individual $i$ under adaptive intervention $(a_1, a_{2R}, a_{2NR})$, where $a_1$, $a_{2R}$, and $a_{2NR}$ represent, respectively, the first stage treatment and the second stage treatment for the responders and non-responders.
Note that in a prototypical SMART, responders continue with their first stage treatment so for ease of notation, we write the complete adaptive intervention as $(a_1, a_{2NR})$ for the remainder of this paper.
Nonetheless, both responders and non-responders belong to this adaptive intervention rather than strictly non-responders.
In a prototypical SMART, each individual has four potential outcomes \citep{rubin1974estimating,holland1986statistics} which correspond to the potential outcomes under each of the four adaptive interventions embedded within the design of the SMART for comparison.

One causal estimand of interest, the difference in mean outcomes between two of the embedded adaptive interventions, may be written as $\E[Y_i^{(a_1,a_{2NR})} - Y_i^{(a_1^*,a_{2NR}^*)}]$, 
where $(a_1,a_{2NR})$ and $(a_1^*,a_{2NR}^*)$ denote two different adaptive interventions embedded within the SMART for comparison.
This causal estimand corresponds to a common primary aim (primary research question) in a SMART: the comparison of two or more of the embedded adaptive interventions \citep{oetting2007statistical}. 
For example, a scientist may want to test the average difference on a continuous, end of study outcome between the most intensive adaptive intervention and the least intensive adaptive intervention.
In the context of ASIC, this could include a comparison of the AI where school professionals receive skills coaching in the first stage and non-responders augment that coaching with a facilitator in the second stage and the AI where school professionals only receive skills coaching regardless of the stage.

Other common primary aims in a SMART include the comparison of first stage treatments ($\E[Y_i^{(a_1,A_{2NR})} - Y_i^{(a_1^*,A_{2NR})}]$) (e.g., the effect of receiving skills coaching in the first stage), or the comparison of second stage treatments among the non-responders ($\E[Y_i^{(A_1,a_{2NR})} - Y_i^{(A_1,a_{2NR}^*)} | \text{Non-Response}]$) (e.g., the effect of augmenting the first stage treatment with a facilitator in the second stage for non-responders).

Recall the purpose of this manuscript is to present easy-to-use strategies for improving efficiency in the estimation of such causal effects.
Before introducing these strategies in the next section, we first review the most basic approach to making this comparison.

\subsection{Primary Aim Analyses in a SMART}
\label{subsec:anal_smart}

Take the prototypical SMART and its embedded adaptive interventions as presented in Table~\ref{tab:embedDTR}.  
Generally, $A_1$ denotes the treatment received during the first stage whereas $A_{2R}$ and $A_{2NR}$ denote the second stage treatment for responders and non-responders respectively.
Thus, each embedded AI may be written as $(a_1, a_{2R}, a_{2NR})$. 
Recall that in a prototypical SMART, we instead write each embedded adapted intervention as $(a_1, a_{2NR})$.
Furthermore, note that it is standard in SMART settings to use contrast coding, i.e., $a_1,a_{2NR} \in \{-1,1\}$. 

\begin{table}[htb]
	\centering
	\begin{tabular}{lccccccc}
		\toprule
		\textbf{AI Label} & \textbf{1st Stage Tx} & \textbf{Resp. Status} & \textbf{2nd Stage Tx} & \textbf{$A_1$} & \textbf{$A_{2NR}$} & \textbf{Cell} & \textbf{d}\\ \midrule
		\multirow{2}{*}{(1,1)}     & B & 1 & B & 1  &    & 1 & \multirow{2}{*}{1}\\
		          & B & 0 & D & 1  & 1  & 2 &   \\
		\multirow{2}{*}{(1,-1)}    & B & 1 & B & 1  &    & 1 & \multirow{2}{*}{2}\\
		          & B & 0 & E & 1  & -1 & 3 &   \\
		\multirow{2}{*}{(-1,1)}    & C & 1 & C & -1 &    & 4 & \multirow{2}{*}{3}\\
		          & C & 0 & D & -1 & 1  & 5 &   \\
		\multirow{2}{*}{(-1,-1)}   & C & 1 & C & -1 &    & 4 & \multirow{2}{*}{4}\\
		          & C & 0 & E & -1 & -1 & 6 &   \\          
		\bottomrule
	\end{tabular}
	\caption{The adaptive interventions embedded within the prototypical SMART presented in Figure~\ref{fig:proto_smart}. $d$ simplifies notation for adaptive intervention labels as discussed in Section~\ref{subsec:anal_smart}}.
	\label{tab:embedDTR}
\end{table} 
Let $\E[Y^{(a_1, a_{2NR})}]$
denote the marginal mean outcome under AI $(a_1, a_{2NR})$
 and $\mu^{(a_1, a_{2NR})} \defeq \E[Y^{(a_1, a_{2NR})}]$.
To simplify notation, we will sometimes replace the four pairs given by $(a_1, a_{2NR})$ with $d = 1, 2, 3, 4$ (see Table~\ref{tab:embedDTR}).
Estimation of $\mu^{(d)}$ is straightforward. 
For a fixed $d$, 
\begin{equation*}
	\hat\mu^{(d)} = \frac{\sum_{i = 1}^n \I^{(d)}(A_{1,i}, R_i, A_{2NR,i})W^{(d)}(A_{1,i}, R_i, A_{2,i}) Y_i}{\sum_{i = 1}^n\I^{(d)}(A_{1,i}, R_i, A_{2NR,i})W^{(d)}(A_{1,i}, R_i, A_{2,i})}, 
\end{equation*}
where $R_i$ is a binary variable denoting response status for $i$, $\I^{(d)}(A_{1,i}, R_i, A_{2NR,i})$ is an indicator denoting that $i$'s observed treatment pathway in the SMART is consistent with adaptive intervention $d$ and $W^{(d)}(A_{1,i}, R_i, A_{2,i})$ represents the weight attached to $i$ \citep{murphy2001marginal}. 
Thus, $\hat\mu^{d}$ is a weighted average of the outcomes consistent with AI $d$. 
These weights correspond to inverse probability of assignment weights \citep{cole2008constructing} and are known because they are a function of the known randomization probabilities. 
In a prototypical SMART, the known weights are $W^{(d)} = 2R + 4(1-R)$. 
These weights are needed to account for the under-representation of non-responders consistent with any given adaptive intervention $d$ \citep{nahum2012experimental}. 
This occurs by design, in that responders are only randomized once whereas non-responders are randomized twice.
To illustrate, see Table~\ref{tab:embedDTR}. 
Cell 1 denotes the responders to $a_1 = 1$, and appears under both AI $(a_1 = 1, a_{2NR} = 1)$ and AI $(a_1 = 1, a_{2NR} = -1)$.
The cells corresponding to non-responders, however, only appear under one AI each.  

An estimator of the covariance between two estimated means is:
\[\widehat{\text{Cov}}(\hat\mu^{(k)}, \hat\mu^{(l)}) = \frac{\sum_{i=1}^n W_i^2 \big(\I_i^{(k)}(Y_i - \hat\mu^{(k)})\big)\big(\I_i^{(l)}(Y_i - \hat\mu^{(l)})\big)}{\sum_{i=1}^n W_i^2 \I_i^{(k)} \I_i^{(l)}}.\]

\noindent For a derivation, see the \href{https://github.com/timlycurgus/SMART-efficiency}{Supplemental Appendix} (also see \citet{nahum2012experimental}). The above formulae can be used to obtain estimates of \textemdash and make statistical inferences about \textemdash the causal effects of one adaptive intervention versus another. 

\section{Techniques to Increase Efficiency when Analyzing SMARTs}
\label{sec:eff_methods}

In this section, we build on the basic estimation approach presented in the previous section by presenting four extensions that have the potential to increase statistical efficiency when analyzing SMARTs.
Many investigators in the educational and behavioral sciences prefer a regression approach to analyzing data from randomized trials. 
Thus, before introducing the four techniques, we now present the approach introduced in Section~\ref{sec:rev} in a regression-based framework that will more easily allow us to adapt our approach to take into account each of the ensuing techniques discussed in this section.

We now consider a marginal structural mean model, i.e., 
$\mu^{(a_1, a_{2NR})}(\boldsymbol\gamma)$
for a prototypical SMART as follows:
\begin{equation}
	\label{eq:msm}
	\mu^{(a_1, a_{2NR})}(\boldsymbol{\gamma}) = \gamma_0 + \gamma_1 a_1 + \gamma_2 a_{2NR} + \gamma_3 a_1 a_{2NR},
\end{equation}
where $a_1,a_{2NR} \in \{-1,1\}$.
Briefly, marginal structural models are a class of models where the parameters are estimated through inverse-probability-of-treatment-weighting.
In this formulation, $\gamma_0$ denotes the intercept and may be interpreted as the grand mean of the observed $Y$.
The other three parameters $(\gamma_1, \gamma_2, \gamma_3)$ can be used to make pairwise comparisons between the four embedded AIs, as well as having their own causal interpretations (see the \href{https://github.com/timlycurgus/SMART-efficiency}{Supplemental Appendix}). 

From Equation~\eqref{eq:msm}, $\mu^{(1,1)}$, the marginal mean for adaptive intervention $(a_1 = 1, a_{2NR} = 1)$ is given by $\gamma_0 + \gamma_1 + \gamma_2 + \gamma_3$ and $\mu^{(-1,-1)}$ is given by  $\gamma_0 - \gamma_1 - \gamma_2 + \gamma_3$. 
Here, the difference in mean outcomes between these two adaptive interventions is given by the linear contrast $(\gamma_0 + \gamma_1 + \gamma_2 + \gamma_3) - (\gamma_0 - \gamma_2 - \gamma_3 + \gamma_4) = 2(\gamma_2 + \gamma_3)$.

We estimate $\boldsymbol{\hat\gamma}$ by solving the following estimating equation \citep{nahum2012experimental}:
\begin{equation}
	\label{eq:esteqn_0}
	0 = \frac{1}{n}  \sum_{i = 1}^{n} \sum_d [\sigma(d)^{-2} \I^{(d)}(A_{1,i}, R_i, A_{2,i}) W^{(d)}(A_{1,i}, R_i, A_{2,i})D^{(d)T}(Y_i - \mu^{(d)}(\boldsymbol\gamma))],
\end{equation}
where $D^{(d)}$ denotes the Jacobian of $\mu^{(d)}(\boldsymbol{\gamma})$ with respect to $\boldsymbol\gamma$ and $\sigma(d)^2$ is a working model for the variance of $Y^{(d)}$. 
For now, we assume that the variance is homogeneous across the various adaptive interventions and thus write $\sigma(d)^2 = \sigma^2$.
Later, this working variance assumption is relaxed.

Note that $\boldsymbol{\hat\gamma}$ is a consistent estimator of $\boldsymbol\gamma$ and its distribution is asymptotically normal. 
For a proof, 
see \citet{lu2016DTR} or \citet{necamp2017comparing}.
Rewrite Equation~\eqref{eq:esteqn_0} as follows:
\begin{equation*}
	\begin{split}
		0 & = \frac{1}{n}  \sum_{i = 1}^{n} \sum_d [\sigma^{-2} \I^{(d)}(A_{1,i}, R_i, A_{2,i}) W^{(d)}(A_{1,i}, R_i, A_{2,i})D^{(d)T}(Y_i - \mu^{(d)}(\boldsymbol\gamma))] \\ &= \frac{1}{n}\sum_{i = 1}^n \mathbf{M}_i(A_{1,i}, R_i, A_{2,i}, Y_i;\boldsymbol{\gamma}),
	\end{split}
\end{equation*}
and define
\begin{equation*}
	\mathbf{B}(\sigma) = \frac{1}{n}  \sum_{i = 1}^{n} \sum_d [\sigma^{-2} \I^{(d)}(A_{1,i}, R_i, A_{2,i}) W^{(d)}(A_{1,i}, R_i, A_{2,i})D^{(d)}D^{(d)T}].
\end{equation*}
An estimate of  Var$(\boldsymbol{\hat\gamma})$ is given by the following plug-in sandwich estimator:
\begin{equation}
	\label{eq:se_0}
	\widehat{\text{Var}}(\boldsymbol{\hat\gamma}) = \frac{1}{n}\Big(\mathbf{\hat B}^{-1}(\hat\sigma)\big(\frac{1}{n}\sum_{i=1}^n \mathbf{\hat M}_i(A_{1,i}, R_i, A_{2,i}, Y_i;\boldsymbol{\hat\gamma}) \mathbf{\hat M}_i^T(A_{1,i}, R_i, A_{2,i}, Y_i;\boldsymbol{\hat\gamma}) \big) \mathbf{\hat B}^{-1}(\hat\sigma)\Big).
\end{equation}

Note that this estimation procedure is a generalized version of the estimation procedure described in Section~\ref{sec:rev}.
Unlike that simplified version, however, this generalized form more easily allows for the adaptations to the method (e.g., incorporating baseline covariates, using repeated measurements, etc.) discussed in the remainder of our paper.

We will now present four techniques to improve efficiency when analyzing SMARTs.
This is not an exhaustive list of methods that may improve efficiency; rather, we have selected four methods that are easily implementable with data and tools that are common to education and behavioral scientists.
Each of the first three adjusts the marginal structural mean model of Equation~\eqref{eq:msm} and the estimating equations in Equation~\eqref{eq:esteqn_0} as necessary. 
The fourth method builds on the third and as such, we will adjust the estimating equations from the third method rather than from the baseline method presented above. 

\subsection{Technique 1: Incorporating Baseline Covariates}
\label{subsec:meth1}

It is widely known that incorporating baseline covariates may increase efficiency in treatment effect comparisons \citep[p.39-41]{bloom2007using} both in education and elsewhere. 
For example, controlling for a pre-test score will often substantially improve precision. 
The gains in efficiency from inclusion of baseline covariates should typically remain present when analyzing a SMART as well. 

Let $\mathbf{X}_i$ denote a vector of mean-centered baseline covariates for individual $i$.
Inclusion of these covariates at baseline necessitates adjustments to the marginal structural mean model in Equation~\eqref{eq:msm}:
\begin{equation}
	\label{eq:msm_meth1}
	\mu^{(a_1, a_{2NR})}(\mathbf{\mathbf{X},\boldsymbol\theta}) = \gamma_0 + \gamma_1 a_1 + \gamma_2 a_{2NR} + \gamma_3 a_1 a_{2NR} + \mathbf{X}\boldsymbol\beta,
\end{equation}
where $\boldsymbol\theta = (\boldsymbol\gamma, \boldsymbol\beta)$. 
We now require a corresponding adjustment to our estimating equation and solve the following:
\begin{equation}
	\label{eq:esteqn_1}
	0 = \frac{1}{n}  \sum_{i = 1}^{n} \sum_d [\sigma^{-2} \I^{(d)}(A_{1,i}, R_i, A_{2,i}) W^{(d)}(A_{1,i}, R_i, A_{2,i})D^{(d)T}(Y_i - \mu^{(d)}(\mathbf{X},\boldsymbol\theta))].
\end{equation}
Intuitively, when $\mathbf{X}$ is correlated with $Y$, we expect the residual errors from applying the marginal model in Equation~\eqref{eq:msm_meth1} to be less variable than the residual errors from applying that in Equation~\eqref{eq:msm}, leading to more efficient estimates. 

\textbf{Remark:} $\mathbf{X}$ may only consist of covariates collected at baseline rather than throughout the course of the study.
Conditioning on variables collected post-baseline may lead to ``collider bias'' \citep{hernan2004structural,cole2010illustrating,elwert2014endogenous} because that covariate may be influenced simultaneously by both the treatment and other known or unknown factors contributing to the outcome. 
As a consequence, controlling for this covariate could lead to spurious associations between the treatment and outcome. 

To illustrate, take ASIC (the SMART introduced in Section~\ref{sec:intro}) which aims to improve CBT delivery provided within a school.
Researchers may choose to collect some school-level metric representing the overall mental health at the school (e.g., the proportion of students with anxiety).
This covariate may be safely accounted for at baseline, but not at subsequent points of the study.
The amount of CBT delivered to students within the school likely affects the overall mental state of those students, yet that metric is also plausibly affected by the intervention aimed at improving CBT delivery. 
Thus, collider bias may arise.
Collider bias could also arise if you inadvertently adjust for response status in the comparison of AIs. 

\subsection{Technique 2: Using Estimated rather than Known Weights}
\label{subsec:meth2}

We discussed the necessity of using a weighted rather than unweighted estimator in Section~\ref{subsec:anal_smart}.
In SMARTs, these weights are known and easy to formulate because we know the randomization probabilities for each participant. 
Thus, responders in a prototypical SMART with equal probability of assignment to each of the treatments receive a weight of 2 and non-responders (who are randomized twice), receive a weight of 4.

While these are the known weights, it may be possible to realize gains in efficiency by estimating the weights instead, i.e. by using $\widehat W$ rather than $W$ \citep{hernan2002estimating,hirano2003efficient,brumback2009note,almirall2014time}. 
The intuition for this is most easily understood as follows. 
The known weights are formulated using the true randomization probabilities $p_{11} = \Prob(A_1 = 1)$ and $p_{21} = \Prob(A_{21} = 1 | A_1, R)$ for the first and second stage assignments.
Nonetheless, for any given SMART with finite sample size $n$, there is likely to be some variation in the proportion of participants assigned to each of the interventions.
The true randomization probability $p_{11}$ will reflect the proportion of participants assigned to $A_1 = 1$ in expectation but for any given SMART, $\hat p_{11}$ will reflect the observed proportion.
This holds for $\hat p_{21}$ as well.

As such, it is possible to estimate weights using the sample probabilities of assignment rather than the known probabilities of assignment (e.g., $\widehat W = (\hat p_{11} \hat p_{21})^{-1}$ rather than $W = (p_{11}p_{21})^{-1}$). 
But researchers need not solely restrict themselves to estimating the randomization probabilities through the sample proportions. 
Instead, there may be further gains to efficiency by modeling the randomization probabilities as a function of other covariates \citep{williamson2014variance}.  
This desire may arise for multiple reasons. 
First, analysts may wish to incorporate information from $\mathbf{X}$ without directly adjusting for $\mathbf{X}$ in the model. 
Alternatively, the researcher may possess some post-baseline auxiliary covariates $\mathbf{L}$ that they cannot control for without the risk of introducing collider bias (see Section~\ref{subsec:meth1}). 
In either of these scenarios, the researcher may implicitly account for variation arising due to $\mathbf{X}$ or $\mathbf{L}$ by using those covariates in their model estimating $\hat p$, the randomization probability. 

With estimated rather than known weights, the marginal structural mean model from Equation~\eqref{eq:msm} remains unchanged.
However, our estimating equation is now:
\begin{equation}
	\label{eq:esteqn_2}
	0 = \frac{1}{n}  \sum_{i = 1}^{n} \sum_d [\sigma^{-2} \I^{(d)}(A_{1,i}, R_i, A_{2,i}) \widehat W^{(d)}(\mathbf{K}_i)D^{(d)T}(Y_i - \mu^{(d)}(\boldsymbol\gamma))],
\end{equation}
where $\mathbf{K}$ denotes the data (if any) used to estimate $\widehat W$. $\mathbf{K}$ can be decomposed into $\mathbf{K}^1$ and $\mathbf{K}^2$ which refer to the variables used to estimate the first and second stage probabilities of assignment.
For example, $\mathbf{K}^1 = (\mathbf{X}, A_1)$ and $\mathbf{K}^2 = (\mathbf{K}^1, R, L, A_2)$ may be used to estimate $\Prob(A_1|\mathbf{X})$  and $\Prob(A_2|\mathbf{X}, A_1, R, L)$ respectively, both of which would be used to estimate $\widehat W$. 
A corresponding adjustment to the standard error is required as well.
For a derivation of the estimated-weight adjusted standard error, see Appendix~\ref{app:wts_asymp}.
In Appendix~\ref{app:add_sims}, we use simulation experiments to examine the effect of estimated weights on efficiency across different sample sizes. 

\subsection{Technique 3: Repeated Measures Analysis with a Working Exchangeable-Homogeneous Variance-Covariance Structure}
\label{subsec:meth3}

As in non-SMART settings, we expect there to be efficiency gains associated with taking advantage of the within-person correlation \citep{ballinger2004using}. 
That is, we expect that obtaining repeated outcome measures data and then applying longitudinal methods could help researchers realize substantial gains in efficiency.
Obtaining repeated measurements should also allow analysts to answer additional secondary research questions like estimating trends in effect sizes over the course of the adaptive intervention.

Take outcome $Y_{it}^{(d)}$, the outcome of individual $i$ in time $t$ when they receive adaptive intervention $d$, where $t = 1, \dots, T$. 
In addition, let $t^*$ denote the time period immediately preceding the second stage randomization. 
One formulation of the marginal structural mean model, proposed in \citet{raudenbush2001comparing,lu2016DTR,seewald2020sample}, is as follows:
\begin{equation}
	\label{eq:msm_3}
	\begin{split}
		\mu_t^{(a_1, a_{2NR})} (\boldsymbol{\gamma}) & = \gamma_0 + \I(t \leq t^*)\big(\gamma_1 t + \gamma_2 a_1 t\big) \\ & + \I(t > t^*)\big(\gamma_1 t^* + \gamma_2 a_1 t^*  + (t - t^*)(\gamma_3 + \gamma_4 a_1 + \gamma_5 a_{2NR} + \gamma_6 a_1 a_{2NR}) \big)
	\end{split}.
\end{equation}

If we allow $t^* = 1$ and we have data collected at $t = 1,2$, then this simplifies to:
\begin{equation*}
	\begin{split}
		\mu_t^{(a_1, a_{2NR})} (\boldsymbol{\gamma}) & = \gamma_0 + \I(t \leq t^*)\big(\gamma_1 + \gamma_2 a_1\big) \\ & + \I(t > t^*)\big(\gamma_1 + \gamma_2 a_1   + \gamma_3 + \gamma_4 a_1 + \gamma_5 a_{2NR} + \gamma_6 a_1 a_{2NR} \big)
	\end{split}.
\end{equation*}

This longitudinal marginal structural model is designed to accommodate  the specific features of the SMART in Figure~\ref{fig:proto_smart}.
For example, the second-stage treatment has not yet occurred at $t = 1$, so $\mu_t$ is merely a function of some intercept $(\gamma_0 + \gamma_1)$ and the first stage treatment $\gamma_2 a_1$. 
By $t = 2$, our marginal structural mean model looks similar to our original marginal structural mean model in Equation~\eqref{eq:msm}, where the intercept in this formulation is now $\gamma_0 + \gamma_1 + \gamma_3$. 

As an example, let us examine ASIC (the SMART introduced in Section~\ref{sec:intro}) once again. 
At the beginning of the study, schools are randomly assigned to one of two treatments. 
Based on their response to these treatments at the end of Stage 1,  non-responders are randomly assigned to one of two augmented treatments at the start of the second stage.
The average weekly number of CBT sessions provided within the school during the final stage of the study is the outcome of interest and used to analyze the various adaptive interventions. 
Measuring average weekly CBT delivered at the end of the earlier phases as well, however, would better allow researchers to precisely isolate the effect of the initial treatment and thus, better estimate the effects of each adaptive intervention as well. 

To estimate $\boldsymbol\gamma$ using a longitudinal approach, we choose to adjust the estimating equation to allow the analyst to specify a working covariance: 
\begin{equation}
	\label{eq:esteqn_3}
	0 = \frac{1}{n}  \sum_{i = 1}^{n} \sum_d [\I^{(d)}(A_{1,i}, R_i, A_{2,i}) W^{(d)}(A_{1,i}, R_i, A_{2,i})D^{(d)T}V(\sigma, \rho)^{-1} (\mathbf{Y}_i - \boldsymbol{\mu}^{(d)}(\boldsymbol\gamma))].
\end{equation}
Here, $\mathbf{Y}_i$ and $\boldsymbol\mu^{(d)}$ are now $(T \times 1)$ column vectors of, respectively, an individual's outcomes and mean outcomes under adaptive intervention $d$ for each time $t$. 
Similarly $D^{(d)}$, the Jacobian of $\boldsymbol\mu^{(d)}(\boldsymbol{\gamma})$ with respect to $\boldsymbol\gamma$, is now a $(T \times p)$ matrix, where $p$ corresponds to the number of parameters $\boldsymbol\gamma$ we are estimating. 

In addition, we now have the option to include a working covariance matrix $V(\sigma, \rho)$ to account for correlation within a participant's outcomes, which may provide a boost to efficiency \citep{vansteelandt2007confounding,tchetgen2012specifying}.
A common choice for the working covariance is the so-called exchangeable-homogeneous working model \citep{seewald2020sample}.  With this working covariance,
the $(T \times T)$ matrix has variance component $\sigma^2$ along the diagonal and $\rho\sigma^2$ along the off-diagonal where $\rho$ denotes the correlation component, i.e. the correlation across time of an individual's outcomes. 
Note that by homogeneous, we mean homogeneity across time and across adaptive intervention.

\textbf{Remark:} There is no requirement that the working covariance model $V^{(d)}(\sigma, \rho)$ be correctly specified; $\boldsymbol{\hat\gamma}$ will remain a consistent estimator of $\boldsymbol\gamma$ regardless of the chosen structure of the covariance matrix \citep{liang1986longitudinal}. 
 Nonetheless, selecting a $V^{(d)}(\sigma, \rho)$ closer to the true covariance matrix should provide greater efficiency.

 \textbf{Remark:} Note that our estimating equation is unbiased due to our proposed weighting structure where the weights for individual $i$ do not change across intervention stages, i.e., $W_{i1} = W_{i2} = \dots = W_{iT}$. 
 To illustrate, non-responders receive a weight of 4 during both the first and second stage even though only one randomization has occurred by the end of the first stage. 
 If instead we use an alternative weighting scheme where the weights directly correspond to the number of randomizations that have taken place by that stage (e.g., weights of 2 and 4 for the two stages respectively), then applying a non-independent working covariance structure will likely introduce bias \citep{boruvka2018assessing}. 
 For a more thorough discussion as well as a proof, see the \href{https://github.com/timlycurgus/SMART-efficiency}{Supplemental Appendix}.

\subsection{Technique 4: Applying a Working Heterogeneous Variance-Covariance Structure}
\label{subsec:meth4}
 
In each of the previous three subsections, we used an analysis method that assumed a constant variance across the four adaptive interventions, i.e., $\sigma(d)^2 = \sigma^2$ where $d$ denotes the adaptive intervention.
In Section~\ref{subsec:meth3}, we additionally assumed the variance was homogeneous across repeated measures, $\sigma_{t_1} = \sigma_{t_2} = \dots = \sigma_T$ for all $d$.  
Nonetheless, even when such working variance assumptions are made, a scientist may choose to analyze the data such that the variance may change over adaptive intervention 
(e.g., $\sigma(d) \neq \sigma(d^*)$), 
over time (e.g., $\sigma_{t_1} \neq \sigma_{t_2}$), or over both adaptive intervention and time.
For the remainder of this section, we restrict ourselves to the scenario where we have repeated measurements and allow for a working exchangeable-heterogeneous variance-covariance structure across the repeated measurements.

Under the constant variance working model, we may decompose $V^{(d)}(\sigma, \rho)$ as follows:
\begin{equation*}
	V^{(d)}(\sigma, \rho) = \sigma^2 \text{Exch}(\rho),
\end{equation*}
where Exch$(\rho)$ is an exchangeable $(T \times T)$ matrix with $1$ along the diagonal and $\rho$ on the off-diagonal.
If we relax the constant variance over repeated measures assumption, we now decompose $V^{(d)}(\boldsymbol\sigma, \rho)$ in the following manner:
\begin{equation*}
	V^{(d)}(\boldsymbol\sigma, \rho) = S(\boldsymbol\sigma) \text{Exch}(\rho) S(\boldsymbol\sigma),
\end{equation*}
where $S(\boldsymbol\sigma)$ is a diagonal $(T \times T)$ matrix with $\sigma_t = \text{Var}^{1/2}(Y_{it})$ along the diagonal. 
This allows us to more precisely estimate the variance of the outcome at each time point, which could lead to gains in statistical efficiency.
For an extreme example of why this may be necessary, take a SMART that examined reading outcomes of students from kindergarten through 12th grade. 
Student performance is generally far noisier for younger students than older students, so assuming a constant variance over these 13 years would be costly; we would be overly confident in our estimates from early years and under-confident in our estimates from later years. 

When we relax this constant working variance assumption, the marginal structural mean model remains the same as in Equation~\eqref{eq:msm_3}, but the estimating equation changes slightly:
\begin{equation}
	\label{eq:esteqn_4}
	0 = \frac{1}{n}  \sum_{i = 1}^{n} \sum_d [\I^{(d)}(A_{1,i}, R_i, A_{2,i}) W^{(d)}(A_{1,i}, R_i, A_{2,i})D^{(d)T}V^{(d)}(\boldsymbol\sigma, \rho)^{-1} (\mathbf{Y}_i - \boldsymbol{\mu}^{(d)}(\boldsymbol\gamma))],
\end{equation}
where $\boldsymbol\sigma$ is $(T \times 1)$ vector $\sigma_{t_1} \dots \sigma_{t_T}$.

\textbf{Remark:} Note that if we did not have repeated measures but wanted to allow for heterogeneity across AIs, then our estimating equation would be identical to that presented in Equation~\eqref{eq:esteqn_0}.

\section{ASIC Results}
\label{sec:ASIC_res}

\subsection{Overview}
\label{subsec:ASIC_overview}

There is evidence to suggest that cognitive behavioral therapy (CBT) can improve learning outcomes among students affected by depressive and anxiety disorders \citep{charvat2012research,zins2004scientific}. 
Even though more children’s mental health is provided in schools than any other child-serving sector, many students do not have access to evidence-based practices such as CBT in schools \citep{martini2012best}. 
Our motivating example is drawn from the Adaptive School-based Implementation of CBT Study (ASIC), a prototypical SMART.  
ASIC’s overarching goal is to develop a three-stage, 44-week, school-level adaptive intervention for overcoming barriers to the adoption and delivery of CBT within high schools in the State of Michigan. 
In ASIC, the duration of stages 1, 2 and 3 are 12, 9 and 23 weeks, respectively. 
The school-level outcome used to illustrate the methods in this paper is the average weekly quantity of CBT delivered at the school, in each stage. 
For purposes of comparing the embedded AIs, the study’s primary endpoint is the average weekly CBT delivered in stage 3.

\subsection{ASIC Study Details}
\label{subsec:ASIC_interventions}

At the beginning of the study, all schools are provided a low-intensity intervention known as Replicating Effectiveness Programs (REP). 
REP includes an easy to understand intervention package with practical guidance on how to implement CBT, day-long didactic training in CBT for all school mental health staff and as-needed, ongoing technical assistance across all three stages of intervention.  
Then at the beginning of stage 1, all schools were randomly assigned (with 50\% probability) to either augment REP with CBT skills Coaching (CST+REP) or not (i.e., continue with REP only). 
At the end of stage 1, a school’s response status is determined:  
A school is identified as “slower responding” (R=0) if the school meets any one of the following two criteria: 
(i) the school failed to provide at least 3 CBT components to $>$10 students during stage 1; or 
(ii) staff report $>$2 barriers to CBT delivery.  
Otherwise, a school is identified as “early responding” (R=1). 
At the beginning of stage 2, slower-responding schools were randomly assigned (with 50\% probability) to augment with Facilitation (FCT) vs. no FCT.  
FCT is an intervention that provides schools with opportunities to discuss barriers to CBT delivery with a ``facilitator'' who regularly meets with leadership and school professionals to help them identify opportunities to overcome barriers.
Early responding schools continue with their current intervention. 
In stage 3, all interventions are discontinued but CBT delivery is still tracked within each school. 

Table~\ref{tab:ASIC_AIs} shows the four interventions embedded in ASIC: two are adaptive (“REP+FCT” and “REP+CST+FCT”) and two are not adaptive (“REP+CST” and “Only REP”). 

\begin{table}[htb]
	\centering
	\begin{tabular}{lrrcrccc}
		\toprule
		\textbf{AI} & \textbf{Lead-In} & \textbf{1st Stage} & \textbf{R} & \textbf{2nd Stage} & \textbf{$A_1$} & \textbf{$A_{2NR}$} & \textbf{Cell} \\ \midrule
		(1,1)   & REP & CST + REP & 1 & CST + REP       & 1  &    & 1 \\
		        & REP & CST + REP & 0 & FCT + CST + REP & 1  & 1  & 2 \\
		(1, -1) & REP & CST + REP & 1 & CST + REP       & 1  &    & 1 \\
		        & REP & CST + REP & 0 & CST + REP       & 1  & -1 & 3 \\
		(-1, 1) & REP & REP       & 1 & REP             & -1 &    & 4 \\
		        & REP & REP       & 0 & FCT + REP       & -1 & 1  & 5 \\
		(-1, -1)& REP & REP       & 1 & REP             & -1 &    & 4 \\
		        & REP & REP       & 0 & REP             & -1 & -1 & 6 \\          
		\bottomrule
	\end{tabular}
	\caption{The interventions embedded within the prototypical SMART known as ASIC, described in Section~\ref{sec:ASIC_res}.}
	\label{tab:ASIC_AIs}
\end{table} 

\subsection{Data Analysis Results}
\label{subsec:ASIC_results}

Prior to data analysis, multiple imputation (40 data sets) was used to impute missing values. 
Standard methods were used for combining estimates, standard errors, and confidence intervals from identical analyses on each of the imputed data sets.

Results using each of the techniques described in Section~\ref{sec:eff_methods} are presented in Table~\ref{tab:ASIC_res}.  
This table provides estimates and 95\% CIs for 
CBT delivery
between the most intensive adaptive intervention (the AI known as “REP+CST+FCT”) and the least intensive intervention (“Only REP” which is not adaptive). 
\begin{table}[htb]
	\centering
	\resizebox{\textwidth}{!}{
	\begin{tabular}{llrccl}
		\toprule
		 & \textbf{Description} & \textbf{Est.} &  \textbf{95\% CI} & \textbf{CI Len.} & \textbf{Citation}
		 \\   \midrule
		\textbf{Tech. 0} & No techniques              & -0.37 & [-2.25, 1.51] & 3.76  &
		\citet{nahum2012experimental}\\
		\textbf{Tech. 1} & Baseline Covariate         & -0.50 & [-2.41, 1.41] & 3.82  &
		\citet{nahum2012experimental} \\
		\textbf{Tech. 2E}& Empirical Weights          & -0.37 & [-2.25, 1.51] & 3.76  &
		\href{https://github.com/timlycurgus/SMART-efficiency}{Supplemental Appendix} \\
		\textbf{Tech. 2M}& Modeled Weights            & -0.75 & [-2.50, 1.00] & 3.49  &
		\href{https://github.com/timlycurgus/SMART-efficiency}{Supplemental Appendix} \\
		\textbf{Tech. 3} & Longitudinal Data          & -0.26 & [-1.98, 1.46] & 3.44  &
		\citet{seewald2020sample} \\
		\textbf{Tech. 4} & Modeled Variance           & -0.01 & [-1.71, 1.69] & 3.40  &
		\citet{seewald2020sample} \\
		\textbf{Ensem. E}  & All four techniques w/ 2E  & -0.28 & [-2.25, 1.69] & 3.93  &
		\href{https://github.com/timlycurgus/SMART-efficiency}{Supplemental Appendix} \\
		\textbf{Ensem. M}  & All four techniques w/ 2M  & -0.60 & [-2.29, 1.09] & 3.37  &
		\href{https://github.com/timlycurgus/SMART-efficiency}{Supplemental Appendix} \\ \bottomrule
	\end{tabular}}
	\caption{ASIC results using each technique for primary outcome analysis, i.e. the difference CBT delivery between the most intensive treatment (REP + CCT + FCT) and the least intensive (REP). Citation provides the source of where to find code to implement these techniques in SMART settings. }
	\label{tab:ASIC_res}
\end{table} 

\begin{figure}[htb]
  \begin{center}
    \includegraphics[scale = 0.45]{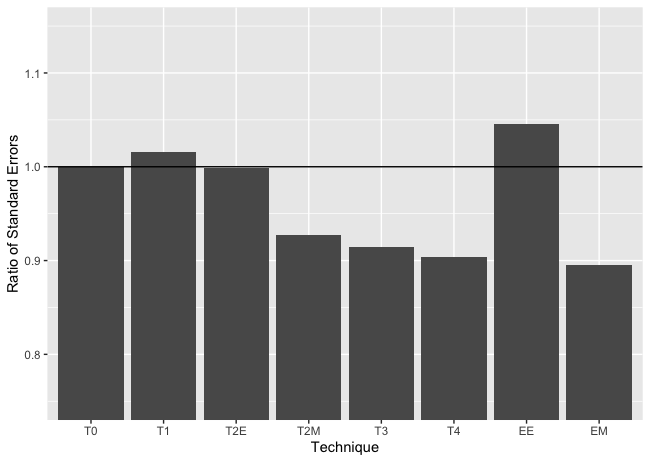}
  \end{center}
  \caption{The ratio of standard errors in relation to the standard error for Technique 0.}
  \label{fig:SE_plot}
\end{figure}

Each technique estimates that less CBT was delivered under the most intensive treatment than under the least intensive treatment, although none of these results are significant. 
We generally find narrower confidence intervals using techniques incorporating longitudinal data as well, suggesting a boost to statistical efficiency.

The full set of pairwise comparisons may be found in Table~\ref{tab:pairwise_comps}. 
These results suggest that Facilitation may improve CBT delivery but that Coaching may be harmful; 
we estimate that each AI with a CST component provides less CBT than the corresponding AI without any coaching.
From these results, it is also clear that the four techniques generally provide smaller confidence intervals than the baseline scenario that adopts none of the approaches. 
The ensemble method with modeled weights is particularly effective and provides the smallest confidence interval for five of the six pairwise comparisons.

\begin{table}[htb]
	\centering
	\resizebox{\textwidth}{!}{
	\begin{tabular}{lrrrrrrrrrrrrrrrr}
		\toprule
		& \multicolumn{8}{c}{\textbf{Diff. in CBT Delivered}} & 
		\multicolumn{8}{c}{\textbf{CI Length}}
		 \\ \cmidrule(lr){2-9} \cmidrule(lr){10-17}
		Design & T0 & T1 & T2E & T2M & T3 & T4 & EE & EM & T0 & T1 & T2E & T2M & T3 & T4 & EE & EM
		 \\   \midrule
		(RCF - R) & -0.37 & -0.50 & -0.37 & -0.75 & -0.26 & -0.01 & -0.28 & -0.60 &
		             3.76 &  3.82 &  3.76 &  3.49 &  3.44 &  3.40 &  3.93 &  \textbf{3.37} \\
		(RF - R)  &  1.37 &  1.41 &  1.37 &  1.04 &  1.31 &  1.33 &  1.30 &  0.84 &
		             5.36 &  5.28 &  5.35 &  4.82 &  4.85 &  4.77 &  4.79 &  \textbf{4.46} \\
		(RC - R)  & -1.05 & -0.92 & -1.05 & -1.28 & -1.06 & -1.06 & -1.28 & -1.41 &
				     3.49 &  3.41 &  3.49 &  3.31 &  3.20 &  \textbf{3.12} &  3.70 &  3.16 \\
		(RC - RF) & -2.24 & -2.33 & -2.42 & -2.32 & -2.37 & -2.39 & -2.58 & -2.25 &
		             5.06 &  4.85 &  5.06 &  4.47 &  4.92 &  4.97 &  5.42 &  \textbf{4.41} \\
		(RCF - RC)&  0.69 &  0.42 &  0.68 &  0.53 &  0.81 &  1.05 &  1.00 &  0.81 &
		             2.76 &  2.80 &  2.76 &  2.35 &  2.57 &  2.70 &  2.73 &  \textbf{2.34} \\
		(RCF - RF)& -1.74 & -1.91 & -1.74 & -1.78 & -1.57 & -1.34 & -1.58 & -1.45 &
		             5.26 &  5.04 &  5.26 &  4.65 &  5.07 &  5.12 &  5.56 &  \textbf{4.49} \\
		\bottomrule
	\end{tabular}}
	\caption{Pairwise comparison of strategies. T2E and T2M refer to Technique 2 with empirical and modeled weights respectively. EE and EM refer to the ensemble method (all four techniques applied at once) using empirical and modeled weights respectively.}
	\label{tab:pairwise_comps}
\end{table} 

\section{Simulations}
\label{sec:sims}

We designed two large simulation studies using modifications of the data generative models presented in \citet{seewald2020sample}.
The purpose of the simulation experiments is to better understand whether, and in what conditions, the four techniques lead to improvements in statistical efficiency. 
We compare each method by itself and all four techniques applied at once. 
We are interested in the efficiency provided by each of the four methods, where we estimate the efficiency in terms of the root mean-squared-error (rMSE), for the comparison of AI $(a_1 = 1, a_{2NR} = 1)$ with AI $(a_1 = -1, a_{2NR} = -1)$. 
We estimate the ``relative efficiency'' between each of the methods and the baseline technique by the ratio of the two rMSEs. 
The first simulation study closely tracks that of \citet{seewald2020sample} and serves as the primary method by which we compare the various techniques.
The second more faithfully mirrors the ASIC SMART discussed in Section~\ref{sec:ASIC_res} and is used to analyze whether the efficiency results from Section~\ref{sec:ASIC_res} likely hold across alternative iterations of ASIC.

\noindent In these studies, we look to answer the following questions:
\begin{itemize}
	\item Under what scenarios, if any, do the four techniques presented in Section~\ref{sec:eff_methods} provide more efficient estimates than the baseline technique that adopts none of the four? 
	\item Are some techniques better or worse than others when we vary the effect size, the within-person correlation $\rho$, and the importance of baseline covariates?
\end{itemize}

We suspect that all four techniques should provide gains in efficiency versus the baseline technique. 
When baseline covariates are more tightly associated with the outcome, we expect Techniques 1 and 2 to perform relatively better than Techniques 3 and 4, because they directly and indirectly account for $\mathbf{X}$.
On the other hand, when $\rho$ increases, we expect Techniques 3 and 4 to provide the most efficient estimates, as they allow for longitudinal data whereas the first two only use the end-of-study outcome. 

\subsection{Data Generative Process}
\label{subsec:data_gen}

Longitudinal data for the first simulation were generated according to the conditional mean model found in Appendix~\ref{app:data_gen}.
The data generative model for the ASIC simulation study is largely similar, but we incorporate multiple covariates $\mathbf{X}$ that correspond to the baseline covariates used in ASIC.
These covariates are generated with the same variance-covariance matrix (see the online \href{https://github.com/timlycurgus/SMART-efficiency}{Supplemental Appendix}) as in the actual study.

In the initial simulation study, 
we set $T = 2$ and $t^*$, the time period immediately preceding the second-stage randomization, equal to one.
For each individual, we generate all potential outcomes under the various adaptive interventions. 
At $t = 1$, we generate $Y_{i1}^{(a_1 = 1)}$ and $Y_{i1}^{(a_1 = -1)}$ for all $i$; at $t = 2$, we must generate four separate potential outcomes where each potential outcome is consistent with one of the four AIs. 
We ``observe'' data by randomly sampling an AI for each individual, where each assignment occurs with equal probability.
We select the potential outcome for each individual that is consistent with their observed AI. 
When analyzing the baseline technique as well as techniques 0-2, we discard outcomes collected prior to the end of the study, i.e. prior to $T = 2$. 

For the ASIC simulation study, $T = 3$ and $t^* = 1$. We generate all potential outcomes under the various adaptive interventions for the two time periods immediately following assignment to a treatment.
Outcomes during the third time period are generated for the AI each individual was assigned to in the previous two time periods.

\subsection{Results}
\label{subsec:results}

We present the relative efficiency for each of the methods, calculated as the ratio of root-mean-squared-errors (RMSE) in Table~\ref{tab:rel_eff}. 
In addition, Table~\ref{tab:rel_eff} presents the percentage of simulations with point estimates closer to the true point estimate than the estimate provided by the baseline method for each technique. 
Formally, let us define $\hat{d}_{k_{j}}$ to be the distance between the estimate using technique $k$ and the parameter for simulation iteration $j$, i.e. $\hat{d}_{k_{j}} = |\hat\beta_{k_j} - \beta|$.
Then Table~\ref{tab:rel_eff} presents $\Prob_j(\hat{d}_{k_{j}} < \hat{d}_{0_{j}})$ for each technique $k$.  

When both $\rho$ and $\nu$ (i.e., Corr$(X,Y)$) are small, the various techniques provide only marginal gains to efficiency;  Technique 1, which directly accounts for baseline covariates, provides the largest benefit at 7\% greater efficiency when $\rho$ and $\nu$ are both set at 0.1.
Nonetheless, this is likely on the lower end of possible values for within-person correlations.

Increasing $\rho$ leads to greater gains in efficiency for each technique, but particularly for Techniques 3, 4, and the ensemble method that uses every technique (denoted EM in Table~\ref{tab:rel_eff}). 
This is unsurprising. 
With low $\rho$, each additional longitudinal observation adds little information so benefits to methods that leverage the longitudinal data are smaller. 
This issue disappears when $\rho$ increases. 
Interestingly, Techniques 1 and 2, which do not use the full, longitudinal data provide slightly greater efficiency gains when $\rho$ increases as well. 

Greater values of $\nu$ generally correspond to greater relative efficiency for each technique in comparison with the baseline scenario as well. 
This benefit, unsurprisingly, is larger for Techniques 1 and 2, the two methods that rely on incorporating information from $\mathbf{X}$ into their estimation procedures. 
When $\rho = 0.8$, increasing $\nu$ from 0.1 to 0.3 improves the relative efficiency of the first two approaches by 8\% and 7\% respectively.  

Comparing across the techniques, it is readily apparent that longitudinal data should be obtained whenever feasible, particularly if $\rho$ is believed to be high. 
In our simulations, techniques incorporating longitudinal data provide slightly greater efficiency than Techniques 1 and 2 when $\rho = 0.3$.
That gap only widens when $\rho$ increases, with the longitudinal approaches providing roughly 50\% greater efficiency when $\rho = 0.8$ and $\nu = 0.3$. 

\begin{table}[htb]
	\centering
	\begin{tabular}{lccccccccccccccc}
		\toprule
		& & & & \multicolumn{5}{c}{\textbf{Relative Efficiency}} & 
		\multicolumn{5}{c}{\textbf{\% Closer than Baseline}}
		 \\ \cmidrule(lr){5-9} \cmidrule(lr){10-14}
		Design & $\delta$ & $\rho$ & $\nu$ & T1 & T2 & T3 & T4 & EM & T1 & T2 & T3 & T4 & EM
		 \\   \midrule
		 \multirow{8}{*}{\textbf{Proto}} & 
		 					  0.3 & 0.1   & 0.1    & 1.07 & 1.04 & 1.02 & 1.02 & 1.04 &
		 					 55.6 & 52.5 & 49.0 & 48.7 & 52.5 \\
		 					 &    &       & 0.3    & 1.08 & 1.06 & 1.03 & 1.04 & 1.06 & 
		 					 53.5 & 53.6 & 51.4 & 50.8 & 54.3  \\
                              &    & 0.3   & 0.1    & 1.08 & 1.06 & 1.08 & 1.08 & 1.09 &
                             54.1 & 53.2 & 53.6 & 54.9 & 53.4   \\
                              &    &       & 0.3    & 1.06 & 1.04 & 1.10 & 1.11 & 1.09 &
                             54.6 & 54.7 & 57.1 & 57.2 & 56.6   \\
                              &    & 0.5   & 0.1    & 1.12 & 1.11 & 1.23 & 1.23 & 1.22 &
                             57.0 & 57.4 & 60.1 & 61.3 & 58.4   \\
                              &    &       & 0.3    & 1.11 & 1.09 & 1.22 & 1.22 & 1.20 &
                             56.3 & 56.0 & 59.6 & 59.4 & 57.5  \\ 
                              &    & 0.8   & 0.1    & 1.11 & 1.12 & 1.69 & 1.68 & 1.66 &
                             57.3 & 56.2 & 67.8 & 67.2 & 66.1   \\
                              &    &       & 0.3    & 1.20 & 1.20 & 1.78 & 1.76 & 1.77 &
                             60.5 & 58.9 & 69.0 & 68.6 & 68.4  \\ 
                             \midrule
                             \multirow{8}{*}{\textbf{ASIC}} & 
                              0.3 & 0.1& \multirow{8}{*}{-}    & 1.00 & 0.96 & 1.09 & 1.09 & 1.04 &
                             48.7 & 45.8 & 58.1 & 57.7 & 51.6   \\
                             &    & 0.3   &     & 1.03 & 0.97 & 1.07 & 1.07 & 1.02 &
                             52.5 & 49.3 & 57.8 & 56.5 & 52.6   \\
                             &    & 0.5   &     & 1.05 & 0.99 & 1.08 & 1.07 & 1.02 & 
                             53.5 & 50.9 & 58.8 & 56.8 & 52.3  \\
                             &    & 0.8   &     & 1.10 & 1.02 & 1.10 & 1.05 & 1.05 & 
                             55.2 & 51.5 & 55.7 & 51.3 & 53.6  \\
                             &0.5 & 0.1   &     & 1.00 & 0.96 & 1.09 & 1.09 & 1.04 &
                             48.7 & 45.8 & 58.0 & 57.7 & 51.6 \\
                             &    & 0.3   &     & 1.03 & 0.97 & 1.07 & 1.07 & 1.02 &
                             52.5 & 49.3 & 57.8 & 56.8 & 52.2 \\
                             &    & 0.5   &     & 1.05 & 0.99 & 1.08 & 1.07 & 1.02 &
                             53.3 & 51.0 & 58.8 & 56.7 & 52.4  \\
                             &    & 0.8   &     & 1.10 & 1.02 & 1.10 & 1.05 & 1.05 &
                             55.2 & 51.3 & 55.7 & 51.2 & 53.4  \\
		\bottomrule
	\end{tabular}
	\caption{Relative efficiency between each technique and the baseline method and the percentage of simulations with a point estimate closer to the true value than that provided by the baseline method. ``EM'' refers to an ensemble method that uses all four techniques together.}
	\label{tab:rel_eff}
\end{table}

Among the longitudinal methods, using the ensemble method with all four techniques at once appears to be marginally worse in terms of efficiency. 
This may be due to the fact that the ensemble method both directly controls for covariates $\mathbf{X}$ and also incorporates $\mathbf{X}$ through the weighting scheme, which wastes degrees of freedom on estimation that did not need to occur.
In Appendix~\ref{app:add_sims}, we look at alternative combinations of techniques to help explain why this occurs. 
In addition, we look at simulations under small and large sample sizes in Table~\ref{tab:change_n} in Appendix~\ref{app:add_sims}. 

The ASIC simulation study shows similar trends both with respect to the general simulation study and to the ASIC results presented in Section~\ref{sec:ASIC_res}.
Technique 1 is marginally more efficient than the standard method whereas Technique 2 shows minimal improvements in efficiency.
Both of these trends correspond well with the confidence interval lengths from the full ASIC results. 
Using the longitudinal data further enhances efficiency.
Unlike the ASIC results, our ensemble method that incorporates all four techniques performs worse than Techniques 3 and 4.
This is a similar trend as what we observed with the general simulation study, although the loss in efficiency relative to Techniques 3 and 4 is greater in the ASIC simulations. 

In sum, we believe that researchers should obtain longitudinal data and incorporate baseline covariates when applicable. 
Improvements in statistical efficiency with a repeated measures outcome analysis are particularly stark when the within-person correlation is expected to be high. 
When the within-person correlation is low, the smaller benefit in efficiency from repeated measurements should be weighed against the additional cost in obtaining those extra measurements. 
Likewise, when the variance is expected to be heterogeneous across either time or adaptive intervention, applying Technique 4 is likely to provide gains to efficiency. 

\section{Discussion}
\label{sec:disc}

Interest in adaptive interventions is increasing both in education practice \citep{raudenbush2020longitudinally} and science \citep{kilbourne2018adaptive,kim2019using,fleury2021early}.
This is unsurprising, as adaptive interventions mirror the sequential and tailored nature of learning within a school.
Education scientists who are engaged in intervention research may have a host of scientific questions about how best to assemble a high-quality AI.
Sometimes, these questions lead to the design of a sequential multiple assignment randomized trial. 
This paper, which was written for an audience of applied statisticians and methodologists in education sciences, both introduces SMARTs and also provides a suite of techniques for their analysis that can be used to enhance statistical efficiency.
These techniques may be particularly important in education settings due to the prevalence of small to moderate effect sizes \citep{kraft2020interpreting}. 
Many of these techniques are common in the analysis of standard randomized trials. 
For example, nearly all randomized trials control for baseline covariates and researchers frequently obtain longitudinal data when that option is available.
The others, applying empirical rather than known weights or allowing for unequal variance across time, however, are less commonly implemented. 

In this paper, we illustrated the application of the various techniques using data from a repeated-measures SMART that aims to develop an AI designed to increase the delivery of CBT across Michigan high schools.
We found that providing REP to all schools followed by providing Facilitation to non-responding schools was the most effective strategy for increasing CBT delivery.
Future work may analyze moderators of effectiveness for Coaching and Facilitation.

We further analyzed and compared the performance of the different proposed techniques using a comprehensive simulation experiment mirroring a prototypical SMART.
Although we limited our focus to the prototypical SMART (with two stages of randomization), the types of efficiency gains observed should generalize to the different types of SMART designs used in practice, even those with three or four randomizations or randomization probabilities different from 50\%.
We find that in general, each of the four techniques proposed in Section~\ref{sec:eff_methods} boosts efficiency in comparison to the baseline method that incorporates none of them. 
However, the magnitude of the efficiency gains varies depending on factors like the within-unit correlation, the sample size, and the correlation between baseline covariates and the outcome.
For example, obtaining a longitudinal outcome and allowing for unequal variances across time can provide far greater enhancements to efficiency than simply using the end-of-study outcome. 
There may be diminishing returns to incorporating each additional technique, yet the ensemble method that uses all four techniques in tandem remains competitive with the others used by themselves. 

We believe that in general, researchers would benefit from obtaining repeated measurements when conducting a SMART.
On top of the likely gains to efficiency, this also allows for analysis of specific aims related to trends in the outcome and to better address potentially negative effects of missing data.
We also recommend obtaining, and controlling for, baseline covariates that are correlated with the outcome of interest. 
Using estimated rather than known weights may be particularly beneficial as well, especially in comparison with their importance in standard randomized trials. 
Modeling the weights allows researchers to incorporate information like response status, a post-baseline measurement, into the weights which implicitly accounts for variation arising due to response status, without introducing collider bias.

In terms of implementation, all of the techniques examined, with the exception of Technique 2, can be performed with standard over the counter statistical software (see Table~\ref{tab:ASIC_res} for citations). 
For example code implementing Technique 2, see the \href{https://github.com/timlycurgus/SMART-efficiency}{Supplemental Appendix}. 

There are a number of interesting directions for future work.
First is whether, and to what extent, the methods presented here generalize to clustered SMARTs \citep{necamp2017comparing}.
Studying this first requires an extension of the longitudinal regression approach that accommodates three levels (e.g., repeated outcome measures, nested within individuals, nested within sequentially randomized clusters) which has not yet been developed.
A particularly interesting statistical question in the clustered context is whether and how to generalize existing finite-sample adjustments when making inferences about the estimated AI effects. 

The second interesting direction is to consider semi-parametric efficient estimators \citep{robins1986new,robins1994correcting} which have the potential to further increase statistical efficiency \citep{robins1995semiparametric,orellana2010dynamic}.
We view this manuscript \textemdash which focuses on methods that are more familiar to applied statisticians in education \textemdash as a first step in this direction.

Third, we found it interesting that the baseline covariate adjustment (Technique 1 in Section~\ref{subsec:meth1}) had largely similar efficiency gains relative to including the baseline covariate in the estimation of the weights 
(Technique 2 in Section~\ref{subsec:meth2}). 
Given this, we conjecture that the latter method will be particularly useful when logit-link marginal models are used to compare AIs on a binary primary outcome \citep{williamson2014variance}.
This would facilitate easier interpretation and statistical inference on both differences in probabilities and log-odds ratios. 

\newpage

\bibliographystyle{apalike}
\bibliography{JEBS_Paper.bib}

\newpage

\begin{appendices}

\section{The Potential Outcome Framework}
\label{app:potential_outcome}

In this appendix, we will review the potential outcomes framework and the assumptions that are necessary to ensure the marginal structural mean $\mu^{(a_1, a_{2NR})}$ is identifiable from the observed data. 
We begin with potential outcomes. 
Let $Y_{it}(a_1, a_{2R}, a_{2NR})$ denote the outcome of individual $i$ in time $t$ if they followed adaptive intervention $(a_1, a_{2R}, a_{2NR})$. 
Note that in the prototypical SMART, those who are responders are not re-randomized to a second stage treatment.
To simplify notation, we will write potential outcome $Y_{it}$ as $Y_{it}(a_1, a_{2NR})$ where $a_{2NR}$ denotes their assignment if they were a non-responder to $a_1$.

In the prototypical SMART, each individual has four potential outcomes: $Y_{it}(1,1)$, $Y_{it}(-1,1)$, $Y_{it}(1,-1)$, and $Y_{it}(-1,-1)$. 
Nonetheless, we never observe all four potential outcomes for each individual; non-responders are only consistent with one adaptive intervention and responders with two. 
Yet under certain assumptions, it is still possible to identify the effect of receiving one AI versus receiving an alternative AI.
These three key assumptions are as follows:
\begin{itemize}
	\item \textbf{Sequential Randomization:} at each stage of a SMART and given the participant's history up to that stage (e.g. $X$, $Y_0$, etc.), the observed treatments $A_1$ and $A_{2NR}$ are assigned independent of any future potential outcomes. That is, $Y_i, R_i \indep A_{1,i}$ and $Y_i \indep A_{2,i} | A_{1,i}, R_i$.
	\item \textbf{Positivity:} $P(A_1 = 1)$ and $P(A_{NR} = 1 | A_1, R = 0)$ are both non-zero. Note that this implies all four probabilities are non-zero. 
	\item \textbf{Consistency:} The observed outcomes, including a participant's response status, are consistent with the potential outcomes under the assigned dynamic treatment regimen.
\end{itemize}

The first two assumptions, sequential randomization and positivity, follow from the design of the SMART. 
Treatment assignment is entirely randomized given the participant's history so the observed treatments occur independently of future potential outcomes. 
Furthermore, randomization of the treatment assignment ensures that each participant will receive each treatment with some probability greater than zero. 

The third assumption requires that the observed outcomes are equivalent to the potential outcomes for the AI assigned to each individual.
This assumption is standard in randomized trials and as such, in SMARTs as well. 

\section{Asymptotics for Proposed Estimators}
\label{app:wts_asymp}

Proofs showing the consistency and asymptotic normality of our proposed estimator may be found in \citet{necamp2017comparing} and \citet{lu2016DTR}. 
In this appendix, we show the asymptotic distribution of the estimator obtained in Section~\ref{subsec:meth2}. 
Namely, we show that under mild conditions, 
\[\sqrt{n}(\hat\beta_{\hat W} - \beta) \rightarrow_p N\big(0, \mathbf{B}^{-1}(\E[\mathbf{MM^T}]- \E[\mathbf{MS}_\omega^T]\E[\mathbf{S}_\omega\mathbf{S}_\omega^T]^{-1}\E[\mathbf{S}_\omega\mathbf{M}^T])\mathbf{B}^{-1}\big),\]
where $\mathbf{B}$ and $\mathbf{M}$ are defined as in Section~\ref{sec:eff_methods} and $\omega$ denotes the parameter in our weight function, i.e. $W(\omega)$, that estimates the probabilities of assignment to the Stage 1 and Stage 2 treatments.
We estimate $\omega$ through a maximum likelihood estimator and use 
$\mathbf{S}_\omega$ to denote the score function of that MLE. 
We assume that $\hat\omega$ is a consistent estimator of $\omega_0$, i.e. $\sqrt{N}(\hat\omega - \omega_0) = O_p(1)$, where $W(\omega_0)$ is the known inverse-probability weight. 
Note that with minor adjustments we can derive the asymptotic distribution of the ensemble estimator that utilizes each of the four techniques. 

Take the following estimating equation corresponding to Equation~\eqref{eq:esteqn_0}, but with $\hat W$ rather than $W$:

\begin{equation*}
	\begin{split}
		0 &= \frac{1}{n}  \sum_{i = 1}^{n} \sum_d [\sigma^{-2} \I^{(d)}(A_{1,i}, R_i, A_{2,i}) \hat W^{(d)}(A_{1,i}, R_i, A_{2,i}, \hat\omega)D^{(d)T}(Y_i - \mu^{(d)}(\boldsymbol\gamma)) \\
		  &=  \frac{1}{n}  \sum_{i = 1}^{n} \mathbf{M}_i(A_{1,i}, R_i, A_{2,i}, Y_i;\hat{\boldsymbol\gamma},\hat{\boldsymbol\omega}).
	\end{split}
\end{equation*}

We now perform a first-order Taylor expansion:
\[0 = \frac{1}{n}\sum_{i=1}^n \mathbf{M}(\cdot) + \frac{1}{n}\sum_{i=1}^n \frac{\partial\mathbf{M}(\cdot)}{\partial \boldsymbol\gamma}(\hat\gamma - \gamma_0) +
\frac{1}{n}\sum_{i=1}^n \frac{\partial\mathbf{M}(\cdot)}{\partial \boldsymbol\omega}(\hat\omega - \omega_0) + o_p(1).\]
Rearranging, we get:
\begin{equation}
	\label{eq:taylor_exp}
\sqrt{n}(\hat\gamma - \gamma_0) = -\E\Big[\frac{\partial \mathbf{M}(\cdot)}{\partial \boldsymbol\gamma}\Big]^{-1}\Big(\frac{1}{\sqrt{n}}\sum_{i=1}^n \mathbf{M}_i + 
\E\Big[\frac{\partial \mathbf{M}(\cdot)}{\partial \boldsymbol\omega}\Big]\sqrt{n}(\hat\omega - \omega_0)\Big) + o_p(1),
\end{equation}
where $\E\Big[\frac{\partial \mathbf{M}(\cdot)}{\partial \boldsymbol\omega}\Big] = \E\Big[\mathbf{M}(\cdot) \mathbf{S}_\omega^T \Big]$.
We now perform another first-order Taylor expansion:
\begin{equation}
	\label{eq:score_exp}
\sqrt{n}(\hat\omega - \omega_0) = -\E\Big[\frac{\partial \mathbf{S}_\omega}{\partial \boldsymbol\omega}\Big]^{-1} \frac{1}{\sqrt{n}}\sum_{i=1}^n \mathbf{S}_{\omega_i} + o_p(1),
\end{equation}
where $-\E\Big[\frac{\partial \mathbf{S}_\omega}{\partial \boldsymbol\omega}\Big]^{-1}$ corresponds to the Fisher information, i.e. $-\E\Big[\mathbf{S}_\omega \mathbf{S}_\omega^T\Big]^{-1}$. 
We then insert Equation~\eqref{eq:score_exp} into Equation~\eqref{eq:taylor_exp}:
\begin{equation*}
\sqrt{n}(\hat\gamma - \gamma_0) = -\E\Big[\frac{\partial \mathbf{M}(\cdot)}{\partial \boldsymbol\gamma}\Big]^{-1}\Big(\frac{1}{\sqrt{n}}\sum_{i=1}^n \mathbf{M}_i - 
\E\Big[\mathbf{M}(\cdot) \mathbf{S}_\omega^T \Big]
\E\Big[\mathbf{S}_\omega \mathbf{S}_\omega^T\Big]^{-1} \mathbf{S}_{\omega_i}\Big) + o_p(1).
\end{equation*}
Thus,
\begin{equation*}
\Var\big(\sqrt{n}(\hat\gamma - \gamma_0)\big) = \mathbf B^{-1}\big(\E[\mathbf{MM}^T] - 
\E[\mathbf{M}\mathbf{S}_\omega^T]
\E[\mathbf{S}_\omega \mathbf{S}_\omega^T]^{-1} \E[\mathbf{S}_{\omega}\mathbf{M}^T]\big)\mathbf B^{-1}.
\end{equation*}

\subsection{Estimation of the Standard Error}

We use plug-in estimators of $\mathbf{B}$ and $\mathbf{M}$ to obtain an estimate of the standard error $(\hat{\boldsymbol{\gamma}}, \hat{\boldsymbol{\omega}})$.

That is,
\[\hat{\mathbf B} = \frac{1}{n} \sum_{i=1}^n \sum_{d}[\hat \sigma^{-2}\I^{(d)}(A_{1,i},R_i,A_{2,i})\hat{\mathbf{W}}^{(d)}(A_{1,i},R_i,A_{2,i};\hat{\boldsymbol\omega})D^{(d)}D^{(d)T}],\]
and
\[\widehat{\mathbf M}^* = \frac{1}{n} \sum_{i=1}^n \hat{\mathbf{M}}_i\hat{\mathbf{M}}_i^T - 
\Big(\frac{1}{n} \sum_{i=1}^n \hat{\mathbf{M}}_i \hat{\mathbf{S}}^T_{\hat{\mathbf\omega}_i}\Big)\Big(\frac{1}{n} \sum_{i=1}^n \hat{\mathbf{S}}_{\hat{\mathbf\omega}_i} \hat{\mathbf{S}}^T_{\hat{\mathbf\omega}_i}\Big)^{-1}\Big(\frac{1}{n} \sum_{i=1}^n \hat{\mathbf{S}}_{\hat{\mathbf\omega}_i} \hat{\mathbf{M}}^T_i\Big),\]
where $\hat{\boldsymbol{S}}_{\mathbf{\omega}_i} = \mathbf{S}_{\hat{\boldsymbol{\omega}}_i}$ and 
\[\hat{\mathbf M}_i = \sum_{d}[\hat \sigma^{-2}\I^{(d)}(A_{1,i},R_i,A_{2,i})\hat{\mathbf{W}}^{(d)}(A_{1,i},R_i,A_{2,i};\hat{\boldsymbol\omega})D^{(d)T}\big(Y_i - \mu(A_{1,i},R_i,A_{2,i};\hat{\boldsymbol{\gamma}}, \hat{\boldsymbol{\omega}})\big)].\]
Then the plug-in estimator for the standard error is $\hat{\mathbf{B}}^{-1}\widehat{\mathbf M}^* \hat{\mathbf{B}}^{-1}$.
For R code to obtain this estimator, see Table~\ref{tab:ASIC_res}.

\section{Additional Simulation Results}
\label{app:add_sims}

First, we examine alternative combinations of techniques rather than simply the ensemble method in Table~\ref{tab:rel_eff_supp}. 
We find, in fact, that approaches that choose between either Technique 1 or Technique 2 (in addition to Techniques 3 and 4), generally outperform the ensemble method that uses all four techniques at once.
This is especially the case for approaches that use known weights but control for a baseline covariate. 
This suggests there are diminishing returns to Techniques 1 and 2. 

\begin{table}[htb]
	\centering
	\resizebox{\textwidth}{!}{
	\begin{tabular}{lccccccccccccccc}
		\toprule
		& & & & \multicolumn{5}{c}{\textbf{Relative Efficiency}} & 
		\multicolumn{5}{c}{\textbf{\% Closer than Baseline}}
		 \\ \cmidrule(lr){5-9} \cmidrule(lr){10-14}
		Design & $n$ & $\rho$ & $\nu$ & T12e & T12m & T134 & T2e34 & T2m34 & T12e & T12m & T134 & T2e34 & T2m34
		 \\   \midrule
		 \multirow{10}{*}{\textbf{Proto}} & 
		 					  250 & 0.1   & 0.1    & 1.05 & 1.06 & 1.05 & 1.03 & 1.01 &
		 					 54.2 & 52.5 & 51.5 & 52.8 & 48.5 \\
		 					 &    &       & 0.3    & 1.06 & 1.07 & 1.06 & 1.06 & 1.03 & 
		 					 54.4 & 53.2 & 50.4 & 53.4 & 50.1  \\
                              &    & 0.3   & 0.1    & 1.07 & 1.07 & 1.10 & 1.08 & 1.07 &
                             53.8 & 53.7 & 55.6 & 53.9 & 54.6   \\
                              &    &       & 0.3    & 1.06 & 1.07 & 1.10 & 1.10 & 1.09 &
                             54.4 & 54.7 & 55.6 & 55.7 & 56.9   \\
                              &    & 0.5   & 0.1    & 1.12 & 1.13 & 1.26 & 1.23 & 1.22 &
                             57.0 & 56.0 & 60.1 & 58.6 & 61.3   \\
                              &    &       & 0.3    & 1.10 & 1.11 & 1.22 & 1.20 & 1.21 &
                             55.5 & 55.6 & 57.2 & 58.3 & 58.6  \\ 
                              &    & 0.8   & 0.1    & 1.13 & 1.13 & 1.69 & 1.66 & 1.66 &
                             56.5 & 56.8 & 67.1 & 66.3 & 66.9   \\
                              &    &       & 0.3    & 1.23 & 1.23 & 1.77 & 1.77 & 1.76 &
                             60.3 & 60.3 & 68.1 & 68.0 & 68.3  \\ \bottomrule
	\end{tabular}}
	\caption{Efficiency performance for alternative combinations of methods. ``e'' refers to methods using estimated empirical weights and ``m'' refers to methods using modeled empirical weights.}
	\label{tab:rel_eff_supp}
\end{table} 

Table~\ref{tab:change_n} shows results under different sample sizes. 
We find that under small sample sizes ($n = 50$), the methods with estimated weights perform relatively worse.
Technique 2 and the ensemble method perform worse on average than the baseline method with $n = 50$ when $\rho$ and $\nu$ are small. 
This aligns with what we expected in Section~\ref{subsec:meth2}.
There is still a benefit to Technique 2 when $n = 1000$, but the gain in efficiency is smaller than when $n = 250$. 

\begin{table}[htb]
	\centering
	\begin{tabular}{lccccccccccccccc}
		\toprule
		& & & & \multicolumn{5}{c}{\textbf{Relative Efficiency}} & 
		\multicolumn{5}{c}{\textbf{\% Closer than Baseline}}
		 \\ \cmidrule(lr){5-9} \cmidrule(lr){10-14}
		Design & $n$ & $\rho$ & $\nu$ & T1 & T2 & T3 & T4 & EM & T1 & T2 & T3 & T4 & EM
		 \\   \midrule
		 \multirow{16}{*}{\textbf{Proto}} & 
		 					  50 & 0.1   & 0.1    & 1.09 & 0.95 & 1.04 & 1.04 & 0.96 &
		 					 58.6 & 51.9 & 54.7 & 54.2 & 49.8 \\
		 					 &    &       & 0.3    & 1.09 & 0.97 & 1.07 & 1.06 & 0.95 & 
		 					 57.5 & 51.7 & 56.9 & 55.7 & 50.0  \\
                              &    & 0.3   & 0.1    & 1.07 & 0.97 & 1.11 & 1.10 & 0.99 &
                             56.5 & 54.1 & 57.2 & 58.9 & 53.6   \\
                              &    &       & 0.3    & 1.12 & 1.00 & 1.17 & 1.16 & 1.04 &
                             58.1 & 52.6 & 59.8 & 58.1 & 54.8   \\
                              &    & 0.5   & 0.1    & 1.11 & 1.04 & 1.25 & 1.23 & 1.15 &
                             59.6 & 54.4 & 60.2 & 61.8 & 58.8   \\
                              &    &       & 0.3    & 1.15 & 1.05 & 1.28 & 1.27 & 1.18 &
                             59.6 & 54.3 & 62.1 & 61.6 & 58.2  \\ 
                              &    & 0.8   & 0.1    & 1.15 & 1.10 & 2.03 & 1.88 & 1.78 &
                             60.1 & 57.8 & 71.2 & 70.7 & 70.8   \\
                              &    &       & 0.3    & 1.27 & 1.15 & 2.06 & 1.98 & 1.83 &
                             62.2 & 59.5 & 74.0 & 73.4 & 71.3 \\ \cmidrule(lr){2-14}
                             &1000& 0.1   & 0.1    & 1.06 & 1.06 & 1.03 & 1.03 & 1.07 &
                             55.1 & 54.3 & 53.7 & 54.2 & 54.6 \\
                             &    &       & 0.3    & 1.05 & 1.05 & 1.02 & 1.02 & 1.04 &
                             53.1 & 55.7 & 50.5 & 50.7 & 53.4  \\ 
                             &    & 0.3   & 0.1    & 1.07 & 1.06 & 1.08 & 1.08 & 1.10 & 55.5 & 56.4 & 57.9 & 58.1 & 57.2  \\
                              &    &       & 0.3    & 1.07 & 1.06 & 1.08 & 1.08 & 1.09 & 
                              52.7 & 52.1 & 58.4 & 59.2 & 56.9  \\
                              &    & 0.5   & 0.1    & 1.07 & 1.06 & 1.15 & 1.15 & 1.17 &
                              56.5 & 56.0 & 58.5 & 58.2 & 57.9 \\
                              &    &       & 0.3    & 1.10 & 1.09 & 1.15 & 1.14 & 1.16 &
                              58.5 & 57.3 & 58.1 & 57.8 & 58.6  \\
                              &    & 0.8   & 0.1    & 1.10 & 1.09 & 1.15 & 1.14 & 1.16 &
                              58.5 & 57.3 & 58.1 & 57.8 & 58.6 \\
                              &    &       & 0.3    & 1.14 & 1.14 & 1.40 & 1.40 & 1.41 &
                              56.0 & 56.5 & 60.0 & 60.4 & 60.9  \\
		\bottomrule
	\end{tabular}
	\caption{Relative efficiency between each technique and the baseline method and the percentage of simulations with a point estimate closer to the true value than that provided by the baseline method. ``EM'' refers to an ensemble method that uses all four techniques together.}
	\label{tab:change_n}
\end{table} 

We also see that our methods, including the ones incorporating repeated measurements with non-independent working variances, are unbiased and have proper confidence interval coverage. 

\begin{table}[htb]
    \centering
    {\resizebox{\textwidth}{!}{
    \begin{tabular}{lccrrrrrrcccccccc}
        \toprule
        & & \multicolumn{6}{c}{\textbf{Bias}} & 
        \multicolumn{6}{c}{\textbf{\% Coverage}}
         \\ \cmidrule(lr){4-9} \cmidrule(lr){10-15}
        Design & $\rho$ & $\nu$ & T0 & T1 & T2 & T3 & T4 & EM & T0 & T1 & T2 & T3 & T4 & EM
         \\   \midrule
         \multirow{8}{*}{\textbf{Proto}}
                        & 0.1   & 0.1    & 0.007 & 0.004 & 0.007 & 0.006 & 0.007 & 0.003 &
                             95.0 & 95.1 & 95.0 & 95.3 & 95.2 & 95.2 \\
                        &       & 0.3    & -0.015 & -0.009 & -0.015 & -0.007 & -0.007 & -0.010 & 
                             94.5 & 95.4 & 95.2 & 94.9 & 94.9 & 94.8  \\
                        & 0.3   & 0.1    & 0.011 & 0.011 & 0.010 & 0.006 & 0.006 & 0.010 & 
                             95.0 & 94.5 & 95.3 & 94.2 & 94.3 & 94.5  \\
                        &       & 0.3    & 0.006 & 0.009 & 0.007 & 0.007 & 0.007 & 0.009 & 
                             95.3 & 95.2 & 95.2 & 95.0 & 95.2 & 95.8   \\
                        & 0.5   & 0.1    & 0.035 & 0.026 & 0.035 & 0.023 & 0.023 & 0.022 & 
                             94.2 & 93.9 & 93.5 & 93.9 & 93.9 & 94.0   \\
                        &       & 0.3    & 0.008 & 0.007 & 0.007 & 0.019 & 0.019 & 0.016 &
                             94.9 & 94.5 & 94.2 & 93.6 & 93.5 & 93.0  \\ 
                        & 0.8   & 0.1    & 0.010 & 0.008 & 0.018 & 0.011 & 0.011 & 0.011 & 
                             95.6 & 95.8 & 96.0 & 96.3 & 96.1 & 96.1   \\
                        &       & 0.3    & 0.012 & 0.004 & 0.011 & 0.015 & 0.014 & 0.012 & 
                             95.3 & 94.9 & 95.6 & 94.9 & 95.4 & 95.3  \\ 
        \bottomrule
    \end{tabular}}}
    \caption{Bias and 95\% confidence interval coverage for each of the methods.}
    \label{tab:bias}
\end{table} 

\section{Simulation Data Generative Process}
\label{app:data_gen}

In this appendix, we will describe the data generative process for our simulation study. 
The levers that will be utilized are within-person correlation $\rho$ and the correlation between baseline covariates and the outcome ($\nu$).
$Y_{t,i}$ denotes the outcome of individual $i$ at stage $t$.

\begin{equation*}
	\begin{split}
		& Y_{0,i} | X_{0,i} = \gamma_0 + \beta_0 X_{0,i} + \epsilon_{0,i} \\
 		& Y_{1,i} | Y_{0,i}, X_{0,i} =  (1 - \rho)(\gamma_0 + \beta_0 X_{0,i}) + \rho Y_{0,i} + \gamma_1 + \gamma_2 a_1 + \epsilon_{1,i}^{a_1} \\
 		& L_i(a_1) | Y_{0,i}, X_{0,i} = g_{a_1}(Y_{0,i}, X_{0,i}) \\
 		& R_i^{a_1} | Y_{0,i}, Y_{1,i}, X_{0,i} = f_{a_1}(Y_{1,i}^{a_1}, X_{0,i}) \\
 		& Y_{2,i} | Y_{0,i}, Y_{1,i}, X_{0,i}, L_{i}(a_1), R_i^{a_1} = 
 		(1 - \frac{2\rho}{1 + \rho})(\gamma_0 + \beta_0 X_{0,i}) + \frac{\rho}{1 + \rho} Y_{0,i} + \frac{\rho}{1 + \rho} Y_{1,i}  \\ 
 		& + (1 - \frac{2\rho}{1 + \rho})(\gamma_1 + \gamma_2 a_1) + \gamma_3 + \gamma_4 a_1 + 
 		\frac{1 - R_i^{a_1}}{1 - r_{a_{1},i}} (\gamma_5 + \gamma_6 a_1)a_{2NR}  \\ 
 		& + (R_i^{a_1} - r_{a_{1},i})(\lambda_1 + \lambda_2 a_1) + \epsilon_{2,i}(R_i^{a_1}, L_i(a_1)),
	\end{split}
\end{equation*}
with $\epsilon_{0,i} \sim N(0, \sigma^2)$, $\epsilon_{1,i} \sim N(0, (1 - \rho^2)\sigma^2)$, and $\epsilon_{2,i}(R_i^{a_1}, L_i(a_1)) \sim N(0, v(R_i^{a_1}) + v(L_i (a_1)))$. 
This should ensure that the variance of $Y_2 | X_0$ is $\sigma^2$.

\end{appendices}

\end{document}